\documentclass[reprint, amsmath,amssymb, aps]{revtex4-2}

\usepackage{graphicx}
\usepackage{subfigure}
\usepackage{multirow}
\usepackage{xcolor, soul}

\usepackage{longtable}
\usepackage[T1]{fontenc}
\usepackage{dcolumn}

\usepackage{graphicx}
\usepackage[colorlinks=true, allcolors=blue]{hyperref}
\usepackage{bm}
\usepackage{amsfonts}  
\usepackage{amsmath}   
\usepackage{amssymb}   
\usepackage{svg}       
\usepackage{blindtext}
\usepackage{subfigure}
\usepackage{multirow}
\usepackage{xcolor, soul}
\usepackage{amsthm}
\usepackage{nicematrix}
\usepackage{MnSymbol}
\usepackage{blindtext}
\usepackage{enumerate}

\theoremstyle{definition}

\begin{document}

\title{Entanglement-efficiency trade-offs in the fusion-based generation of photonic GHZ-like states}
\author{A.\,A.\,Melkozerov$^{1,2}$}
\email{melkozerov.aa17@physics.msu.ru}
\author{M.\,Yu.\,Saygin$^{3,2}$}
\author{S.\,S.\,Straupe$^{3,1,2}$}

\affiliation {\it $^1$ Russian Quantum Center, Bolshoy Boulevard 30, building 1, Moscow, 121205, Russia \\
\it $^2$ Quantum Technology Center, M.\,V. Lomonosov Moscow State University, Leninskie Gory 1, Moscow, 119991, Russia \\
\it $^3$ Sber Quantum Technology Center, Kutuzovski prospect 32, Moscow, 121170, Russia
}

\date{July, 2025}

\begin{abstract} 

Probabilistic entangling measurements are key operations in linear-optical quantum technologies, enabling the generation and manipulation of high-dimensional quantum states. While prior research has focused predominantly on specific entangled states, notably graph states and Greenberger-Horne-Zeilinger (GHZ) states, broader classes of states with variable entanglement remain underexplored. In this work, we present a linear-optical approach for generating and fusing GHZ-like states, which generalize standard GHZ states to include variable entanglement degrees. We introduce two schemes based on modified fusion gates that allow flexible control over generation efficiency and the entanglement of the output states. These results offer a promising pathway toward resource-efficient entangled-state generation for scalable quantum computing and communication.

\end{abstract}

\maketitle 

\section{Introduction}

The creation and manipulation of high-dimensional entangled states are crucial for many photonic quantum technologies, including multiparty quantum networks \cite{Farouk2017, Wang2006, Gao2005, Tth2014, Hyllus2012} and quantum computing \cite{GimenoSegovia2015, Bartolucci2023, Grassl}. Recent linear-optical approaches, such as the fusion-based quantum computing paradigm, rely on the application of entangling measurements to pre-generated entangled states~\cite{Bartolucci2023}.

However, the generation of photonic entanglement in dual-rail encoding is inherently probabilistic. Modern methods exploit single-photon interference in multimode linear-optical circuits and photonic measurements to prepare and manipulate entanglement~\cite{Bartolucci2021, Forbes2025, Carolan2015, Varnava2008}. Although experimentally feasible, these approaches suffer from low success probabilities, resulting in significant technical overhead. This motivates the search for more efficient strategies for linear-optical entanglement generation.

Most versatile linear-optical protocols involve the probabilistic generation of small resource states, followed by the application of entangling fusion measurements to create large-scale states \cite{Zaidi2015, Bartolucci2021, Browne2005, Kieling2007, GimenoSegovia2015, https://doi.org/10.48550/arxiv.2410.20261}. The capabilities and limitations of standard fusion-based methods have been studied in the context of generating stabilizer states \cite{Aaronson2004}, in particular, multiqubit Greenberger-Horne-Zeilinger (GHZ) and graph states \cite{Hein2004}, which are key components in well-established quantum computing models and quantum networks~\cite{MergingQuantumRepeaters}. Notably, without additional resources, such as extra photons beyond those encoding the qubits, the success probability of fusing two maximally entangled qubit states cannot exceed 50\% \cite{Calsamiglia2001, https://doi.org/10.48550/arxiv.2406.15666}. This probability can be increased using extra separable \cite{Ewert2014, https://doi.org/10.48550/arxiv.2410.20261} or entangled \cite{Grice2011, Ewert2014, https://doi.org/10.48550/arxiv.2410.20261} photons.

However, recent investigations in quantum computing introduce novel formalisms that represent wider ranges of states and codespaces \cite{Gross2007, Ni2015, Webster2022}, motivating the need for new linear-optical methods to generate such states. In this paper, we focus on the linear-optical generation of multiqubit GHZ-like states, a generalization of the standard GHZ states that exhibit varying degrees of entanglement. These states are useful in algorithms such as quantum teleportation \cite{Jeong2006, Modawska2008, PATHAK2011}, and were recently proven to be locally unitarily equivalent to weighted hypergraph states \cite{https://doi.org/10.48550/arxiv.2408.02740}, the generalization of graph states. Weighted graph states, in turn, find various applications in quantum algorithms \cite{Hartmann2007, Plato2008, Anders2007}, including measurement-based quantum computing \cite{Gross2007}. Moreover, these states are encompassed by the recent extension of Pauli stabilizer formalism \cite{Webster2022}, a powerful tool underpinning many modern quantum error correction \cite{Grassl} and quantum computing \cite{Bartolucci2023} protocols.

We propose an approach for creating multiqubit GHZ-like states using well-known linear-optical fusion gates and analyze its performance. Our method involves sequential application of probabilistic entangling measurements to small, pre-generated GHZ-like resource states, fusing them into larger states. We show that:
\begin{itemize}
\item The success probabilities for fusing GHZ-like states can surpass the 50\% limit associated with Pauli stabilizer states.
\item Our modifications of fusion gates enable control over the degree of entanglement of the resulting states.
\item Large GHZ-like states can be generated with significantly improved efficiency, at the cost of entanglement degree.
\end{itemize}

We introduce two protocols. The first one begins with the creation of several arbitrarily entangled small resource states and applies a sequence of fusion gates on them to construct the target state. The second algorithm uses specific resource states but achieves higher efficiency. We quantitatively compare both protocols with existing approaches and explore the trade-off between entanglement and success probability. Our estimations show that the probability of state generation can be substantially increased at the cost of reducing the entanglement degree of the target states.

The paper is organized as follows. Section~\ref{sec:preliminaries} introduces the target states and provides an overview of the generation approach. Section~\ref{sec:fusion_gates} details the operation of standard linear-optical fusion gates, while Section~\ref{sec:mod_f_g} investigates the properties of the modified fusion gate. We propose schemes for generating GHZ-like states in Section~\ref{sec:state_gen}.  Section~\ref{sec:performance} evaluates their performance compared to existing methods for creating maximally entangled GHZ states. Finally, Section~\ref{sec:conclusion} discusses the results and directions for future research.

\section{Linear-optical GHZ-like states}\label{sec:preliminaries}

We consider the linear-optical generation of the following $n$-qubit states:
    \begin{equation}\label{eqn:n_GHZ}
        |G_{n}(\alpha)\rangle = \text{cos}(\alpha) \left|0 \right\rangle^{\otimes{}n} + \text{sin}(\alpha)\left|1\right\rangle^{\otimes{}n},
    \end{equation}
where $\left|0 \right\rangle$ and $\left|1 \right\rangle$ form a logical basis of the qubit. The parameter $0\le\alpha\le{}\pi/4$ is a Schmidt angle that quantifies the degree of entanglement of the state~\cite{Bengtsson2006}. Since the parameter values $\alpha=0$ and $\alpha=\pi/4$ correspond to separable and maximally entangled GHZ states, respectively, we call Eq.~\eqref{eqn:n_GHZ} a GHZ-like state.

We investigate the linear-optical generation of the dual-rail encoded states, since this encoding method is the most convenient and widely used in photonic quantum computing~\cite{gimeno2016}. Logical states are represented by a single photon occupying either of two optical modes: $|0\rangle = |10\rrangle = \hat{a}_1^{\dagger}|vac\rrangle$ and $|1\rangle = |01\rrangle = \hat{a}_2^{\dagger}|vac\rrangle$, where $\hat{a}_1^{\dagger}$ and $\hat{a}_2^{\dagger}$ are the creation operators acting on the first and second optical modes, respectively, and $|vac\rrangle$ is the vacuum state $|0\rrangle^{\otimes{}n}$. We designate the physical Fock states by double ket brackets to distinguish them from the logical qubit states.

\begin{figure}[ht]
    \includegraphics[width=3.3in]{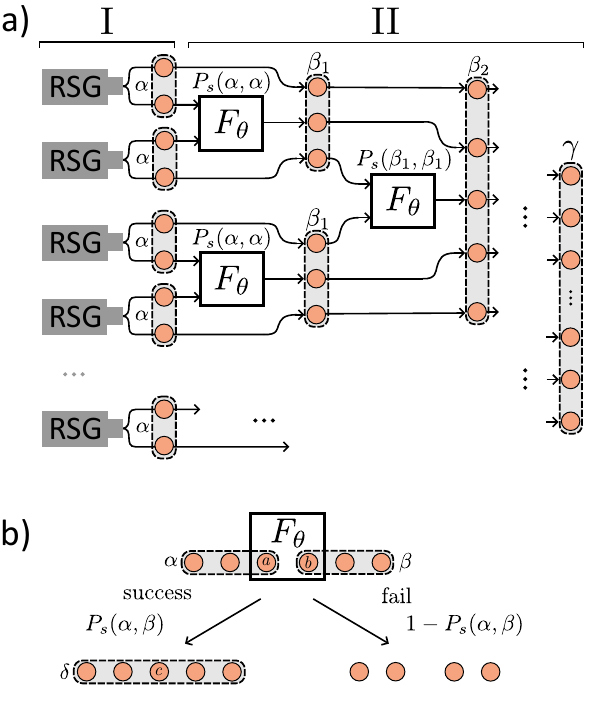}
    \caption{\textbf{Linear-optical generation of multiqubit GHZ-like states.} a) The scheme employs sequential probabilistic entangling fusion operations applied to constant-size resource states $|G_{r}(\alpha)\rangle$ produced by resource state generators (RSGs). Each N-qubit state is represented as N connected nodes labeled with the corresponding Schmidt angles.
    b) The fusion success probability depends on the Schmidt angles $\alpha$ and $\beta$ of the fused states. The Schmidt angle $\delta$ of the resulting state can be controlled via the fusion gate parameter $\theta$.}
    \label{fig:idea}
\end{figure}

Fig.~\ref{fig:idea} illustrates the generation of GHZ-like states of the form \eqref{eqn:n_GHZ} according to our proposal. It consists of two stages: in the first, the \textit{resource state generation stage}, multiple small, constant-sized resource states $|G_{r}(\alpha)\rangle$ are generated (part I of the scheme in Fig.~\ref{fig:idea}a). In the \textit{fusion stage}, denoted as part II in Fig.~\ref{fig:idea}a), these smaller states are fused into larger ones, creating the desired target state as a result.

Similar fusion-based methods are well known for generating maximally entangled states~\cite{gimeno2016, Bartolucci2021}. Typically,  $2$- or $3$-qubit resource states are considered, depending on the particular scheme and the type of gates used, so that they can be generated relatively easily. In particular, the resource states can be created using heralded non-deterministic schemes, such as those proposed for maximally-entangled~\cite{Forbes2025, Stanisic2017, Gubarev2020,Carolan2015, Bartolucci2021, Fldzhyan2021, Bartolucci2021, Varnava2008, Bhatti2024} and non-maximally entangled states~\cite{Fldzhyan2023}. They can also be generated (near)deterministically using switching networks~\cite{switch_networks}. Resource state generation methods can be extended to the class of non-maximally entangled states considered in our schemes. In particular, such $2$-qubit states $|G_{2}(\alpha)\rangle$ have been theoretically investigated in~\cite{Fldzhyan2023} and experimentally demonstrated in~\cite{Skryabin2024}.

In this work, we focus on the fusion stage, which involves a sequence of entangling fusion operations to obtain a GHZ-like target multiqubit state $|G_{N}(\gamma)\rangle$ as the final result. In the schemes, each fusion gate is applied to two GHZ-like states $|G_n(\alpha)\rangle$ and $|G_m(\beta)\rangle$. If successful, the states are fused into a larger GHZ-like state $|G_{(k)}(\delta)\rangle$ at the cost of destroying $1$ or $2$ qubits depending on the type of fusion gate. As will be shown in Sec.~\ref{sec:fusion_gates}, the fusion success probability $P_s$ depends on the entanglement degrees $\alpha$ and $\beta$ of the initial states. Importantly, the fusion success probability can exceed the limit of $1/2$ associated with the fusion of maximally entangled states. Furthermore, the degree of entanglement of the fused states can be controlled using specially designed linear-optical circuits for the fusion gates, which we propose in Fig.~\ref{fig:fusion_gates}.

By sequentially executing appropriate fusion operations, we obtain progressively larger states, eventually creating a target $N$-qubit GHZ-like state $|G_N(\gamma)\rangle$. However, it should be noted that the failure of a single fusion operation leaves the qubits of the fused states unentangled, requiring a restart of the fusion sequence. Hence, the success probability of the scheme equals the product of the respective fusion success probabilities:
\begin{equation}\label{eqn:pproduct}
        P_{\text{gen}}=\prod{P_s},
\end{equation}
where the product is taken over all the fusion operations in the scheme.  

\section{Fusion with standard gates}\label{sec:fusion_gates}

Standard type-I and type-II fusion gates can probabilistically entangle two initially separated qubit states \cite{Browne2005}. Their action can be described as positive operator-valued measures (POVMs) corresponding to measurements of Pauli stabilizer operators \cite{gimeno2016}. For the analysis of subsequent protocols, however, the Hilbert-space representation offers a more intuitive understanding.

\begin{figure}[t]
    \centering
    \includegraphics[width=2.5in]{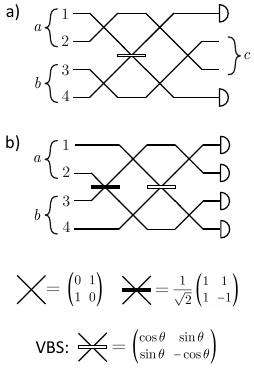}
    \caption{\textbf{Linear-optical circuits realizing modified fusion gates}: a) Circuit implementing the modified type-I fusion gate; b) circuit implementing the modified type-II fusion gate. Both gates act on two input qubits $a$ and $b$ from two initially separated states. The specific fusion operation is determined by the splitting ratio of the variable beam-splitter (VBS), defined by the parameter $\theta$. Standard fusion gates correspond to the special case $\theta = \pi/4$.}
    \label{fig:fusion_gates}
\end{figure}

Fig.~\ref{fig:fusion_gates} shows optical circuits for modified fusion gates incorporating variable beam splitters $\text{VBS}(\theta)$, where the splitting ratio is parametrized by an angle $\theta$, with $0 \le \theta \le \pi/4$. In this section, we focus on standard fusion gates, corresponding to the case $\theta = \pi/4$. These gates consist of 50:50 beam splitters, described by $2\times2$ Hadamard matrices, along with mode swaps and photon-number-resolving detectors (PNRDs).

\subsection{Type-I fusion}

The circuit for the standard type-I fusion gate is obtained from the general setup shown in Fig.~\ref{fig:fusion_gates}(a) by setting $\theta = \pi/4$. The initial state, consisting of two distinct GHZ-like states, takes the form:
\begin{equation}\label{eqn:input_state}
    \begin{aligned}
    |\psi^{(in)} \rangle&= |G_n(\alpha)\rangle|G_m(\beta)\rangle = \\
    &=\left(\text{cos}(\alpha) |0\rangle_a |\overline{0}\rangle_{A} + \text{sin}(\alpha) |1\rangle_a|\overline{1}\rangle_{A}\right) \otimes \\
    &\otimes \left(\text{cos}(\beta) |0\rangle_b |\overline{0}\rangle_{B} + \text{sin}(\beta) |1\rangle_b|\overline{1}\rangle_{B}\right),
\end{aligned}
\end{equation}
Here, we single out qubits $a$ and $b$ that enter the fusion gate, and use $|\overline{0}\rangle_A = |0\rangle^{\otimes(n-1)}$ and $|\overline{1}\rangle_A = |1\rangle^{\otimes(n-1)}$ as shorthand for the remaining qubits. In dual-rail encoding, this state can be expressed in terms of photon creation and annihilation operators as:
\begin{equation}\label{eqn:in_state_second_quantization}
    \begin{aligned}
        &| \psi^{(in)}\rangle = \left( \hat{a}^{\dagger}_1 \hat{A}_0 + \hat{a}^{\dagger}_2 \hat{A}_1  \right) \otimes \left( \hat{a}^{\dagger}_3 \hat{B}_0 + \hat{a}^{\dagger}_4 \hat{B}_1  \right) |vac\rrangle,
    \end{aligned}
\end{equation}
where $\hat{A}_0|vac\rrangle = \cos({\alpha}) |\overline{0}\rangle_{A}$, $\hat{A}_1|vac\rrangle = \sin({\alpha}) |\overline{1}\rangle_{A}$, $\hat{a}^{\dagger}_1 |vac\rrangle=|0\rangle_a$, $\hat{a}^{\dagger}_2 |vac\rrangle=|1\rangle_a$, and similarly for the second state $\hat{B}_0|vac\rrangle = \cos({\beta}) |\overline{0}\rangle_{B}$, $\hat{B}_1|vac\rrangle = \sin({\beta}) |\overline{1}\rangle_{B}$, $\hat{a}^{\dagger}_3 |vac\rrangle=|0\rangle_b$, $\hat{a}^{\dagger}_4 |vac\rrangle=|1\rangle_b$.

The unitary $4\times{}4$ matrix of the circuit $\hat{\mathcal{U}}_{fus}^{(I)}$ defines the transformation of the dual-rail encoded initial states. Constructing it from the matrices of the individual optical elements in Fig.~\ref{fig:fusion_gates}(a) with $\text{VBS}(\theta=\pi/4)$, we obtain:
\begin{equation}\label{Type_1_res}
    \begin{aligned}
        \hat{\mathcal{U}}_{fus}^{(I)}|\psi^{(in)}\rangle &= \Bigl(\frac{1}{2}\left( (\hat{a}^{\dagger}_1)^2 -(\hat{a}^{\dagger}_4)^2 \right) \hat{A}_0 \hat{B}_1 + \hat{a}^{\dagger}_2 \hat{a}^{\dagger}_3 \hat{A}_1 \hat{B}_0 +   \\
        &+\frac{1}{\sqrt{2}} \hat{a}^{\dagger}_1 \left( \hat{a}^{\dagger}_2 \hat{A}_0 \hat{B}_0 + \hat{a}^{\dagger}_3 \hat{A}_1 \hat{B}_1 \right) + \\
        & +\frac{1}{\sqrt{2}} \hat{a}^{\dagger}_4 \left( \hat{a}^{\dagger}_2 \hat{A}_0 \hat{B}_0 - \hat{a}^{\dagger}_3 \hat{A}_1 \hat{B}_1 \right)\Bigr)|vac\rrangle  .
    \end{aligned}
\end{equation}
The state resulting from the measurement of modes 1 and 4 depends on the specific photodetection outcome. Only those outcomes in which a single photon is detected in either mode 1 or 4 lead to a successful event. In such cases, an entangling measurement is applied, and the resulting state takes the form $\left(\hat{a}^{\dagger}_2 \hat{A}_0 \hat{B}_0 \pm \hat{a}^{\dagger}_3 \hat{A}_1 \hat{B}_1\right)/\sqrt{P_{s_1}} \; |vac\rrangle$, where $P_{s_1}$ is the success probability. The explicit form of the state is:
    \begin{equation}\label{Type_1_suc}
        \begin{split}
            |\psi^{(out)}\rangle = \frac{1}{\sqrt{P_{s_1}}}&\left(\text{cos}(\alpha) \text{cos}(\beta) |0\rangle_c |\overline{0}\rangle_{A}|\overline{0}\rangle_{B} \pm \right.\\
            &\left. \pm \text{sin}(\alpha) \text{sin}(\beta) |1\rangle_c |\overline{1}\rangle_{A} |\overline{1}\rangle_{B}\right), 
        \end{split}
    \end{equation}
assuming $\hat{a}^{\dagger}_2 |vac\rrangle = |0\rangle_c$, $\hat{a}^{\dagger}_3 |vac\rrangle = |1\rangle_c$ for the leftover qubit. 

Two initially separated states are fused together into $k$-qubit ($k=n+m-1$) GHZ-like state:
    \begin{equation}
        |G_{k} (\delta)\rangle  =\text{cos}(\delta) |\overline{0}\rangle + \text{sin}(\delta) |\overline{1}\rangle,
    \end{equation}
where the Schmidt angle $\delta$ of the state is connected to the Schmidt angles of the initial states through the relation: 
    \begin{equation}\label{eqn:param_connection}
        \tan(\delta) = \pm\tan(\alpha)\tan(\beta)
    \end{equation}
where the sign depends on a particular detection pattern.

\begin{figure}[t]
        \includegraphics[width=3.3in]{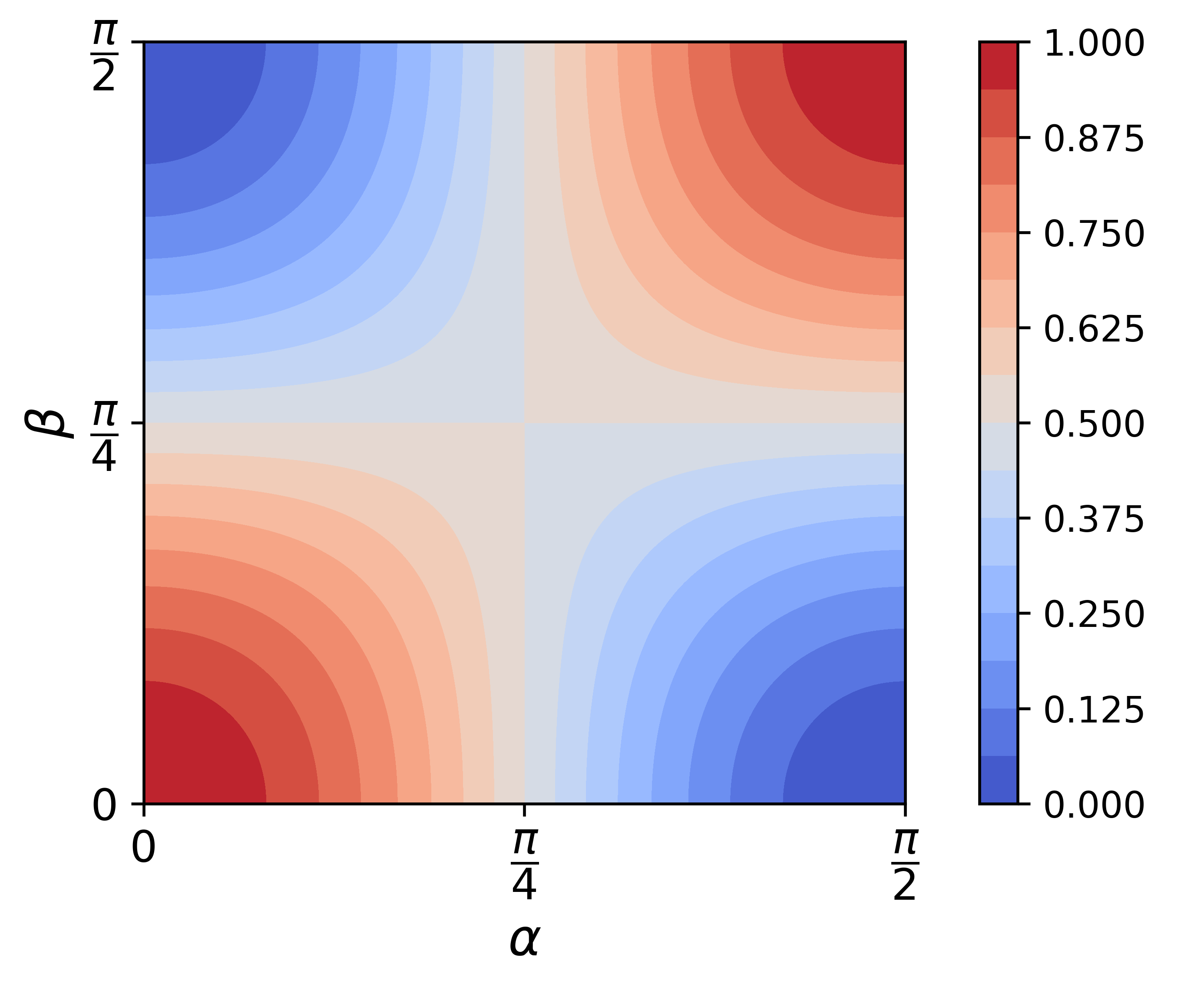}
        \caption{\textbf{Fusion success probability as a function of the degrees of entanglement of the input states}: the trade-off between the success probability of the standard type-I and type-II fusion gates and the Schmidt angles $\alpha$ and $\beta$ of the input states (\ref{Type_1_prob}).}
        \label{fig:succ_prob}
    \end{figure}

Success probabilities of the corresponding patterns are calculated from \eqref{Type_1_res}:
    \begin{equation}
        P_{\pm}=\frac{1}{2}\bigl(\cos^2(\alpha)\cos^2(\beta) + \sin^2(\alpha)\sin^2(\beta) \bigr).
    \end{equation}
The total success probability is depicted in Fig.~\ref{fig:succ_prob}:
    \begin{equation}\label{Type_1_prob}
        P_{s_1}=P_{+} + P_{-}=\frac{1}{2}\bigl( 1+\text{cos}(2\alpha)\text{cos}(2\beta) \bigr).
    \end{equation}

The fusion fails when both input photons or no photons are measured. In this case, the resultant state is $\hat{A}_0 \hat{B}_1 |vac\rrangle$ or $\hat{A}_1 \hat{B}_0 |vac\rrangle$.

\subsection{Type-II fusion}\label{sec:type2fusion}

The standard type-II fusion circuit corresponds to the setup shown in Fig.\ref{fig:fusion_gates}(b) with $\theta=\pi/4$. This circuit transforms the initial state \eqref{eqn:in_state_second_quantization} into:
\begin{equation}\label{Type_2_res}
    \begin{aligned}
        &\hat{\mathcal{U}}^{(II)}_{fus}|\psi^{(in)}\rangle =\frac{1}{2} \Bigl( \hat{a}^{\dagger}_1 \hat{a}^{\dagger}_2 \left(\hat{A}_0 \hat{B}_0 + \hat{A}_1 \hat{B}_1 \right) - \hat{a}^{\dagger}_3 \hat{a}^{\dagger}_4 \left(\hat{A}_0 \hat{B}_0 + \hat{A}_1 \hat{B}_1 \right) + \\
        & +\hat{a}^{\dagger}_2 \hat{a}^{\dagger}_4 \left(\hat{A}_0 \hat{B}_0 - \hat{A}_1 \hat{B}_1 \right) - \hat{a}^{\dagger}_1 \hat{a}^{\dagger}_3 \left(\hat{A}_0 \hat{B}_0 - \hat{A}_1 \hat{B}_1 \right) + \\
        & + \left((\hat{a}^{\dagger}_2)^2 - (\hat{a}^{\dagger}_3)^2\right) \hat{A}_1 \hat{B}_0 + \left((\hat{a}^{\dagger}_1)^2 - (\hat{a}^{\dagger}_4)^2\right) \hat{A}_0 \hat{B}_1 \Bigr)|vac\rrangle.
    \end{aligned}
\end{equation}
The gate succeeds when the detectors register two photons in different modes. In this case, the final state is $\left( \hat{A}_0 \hat{B}_0 \pm \hat{A}_1 \hat{B}_1\right)/\sqrt{P_{s_1}} |vac\rrangle$, or equivalently:
\begin{equation}\label{Type_2_suc}
    \begin{split}
        |\psi^{(out)}\rangle= \frac{\text{cos}(\alpha) \text{cos}(\beta) |\overline{0}\rangle_{A}|\overline{0}\rangle_{B} \pm \text{sin}(\alpha) \text{sin}(\beta) |\overline{1}\rangle_{A} |\overline{1}\rangle_{B}}{\sqrt{P_{s_1}}}, 
    \end{split}
\end{equation}
where $P_{s_1}$ is the total success probability. This matches the success probability given in Eq.~\eqref{Type_1_prob} for the type-I fusion gate.

Fig.~\ref{fig:succ_prob} illustrates the success probability $P_{s_1}$, defined by Eq.~\eqref{Type_1_prob} as a function of the entanglement parameters of the initial states. For $\alpha = \beta = \pi/4$ the success probability equals $1/2$, which corresponds to the known upper bound for fusing maximally entangled states without auxiliary photons or modes. However, the success probability increases as the degrees of entanglement of the fused states decrease, reaching a maximum of $1$ in the limit $\alpha,\beta \rightarrow 0$. 

The success probability of fusing maximally-entangled states can be boosted by employing additional photonic resources \cite{Grice2011, Ewert2014, https://doi.org/10.48550/arxiv.2410.20261}. Similarly, the fusion of GHZ-like states can be boosted using the same auxiliary resource states. For instance, using four additional single photons \cite{Ewert2014} or a Bell pair \cite{Grice2011}, the success probability of the fusion (\ref{Type_1_prob}) can be increased to 

\begin{equation}\label{Type_1_boosted}
    P_{b}=\frac{3}{4} + \frac{1}{4}\text{cos}(2\alpha)\text{cos}(2\beta).
\end{equation}

\section{Fusion with modified gates} \label{sec:mod_f_g}

\begin{figure}[h]
    \includegraphics[width=2in]{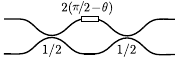}
    \caption{\textbf{Integrated photonic implementation of a variable beam splitter} using two 50:50 beam splitters and a phase shift.}
    \label{fig:vbs}
\end{figure}

When fusing GHZ-like states using standard fusion gates, the resulting GHZ-like states have Schmidt angles determined by the fixed relation~\eqref{eqn:param_connection}, leaving no flexibility to tune the entanglement of the output states. To overcome this limitation, we propose linear optical circuits that incorporate a variable beam splitter (VBS), as shown in Fig.~\ref{fig:fusion_gates}. These circuits enable adjustable control over the degree of entanglement in the output states. Structurally, they are similar to standard fusion gates, but the static balanced beam splitter is replaced with a tunable one, characterized by the following transfer matrix:
    \begin{equation}
    \text{VBS}(\theta) =\left(
    \begin{matrix}
        \cos{\theta} & \sin{\theta}\\
        \sin{\theta} & -\cos{\theta}
    \end{matrix}
    \right),
    \end{equation}
where the angle parameter $\theta$ parametrizes the element. A reconfigurable VBS of this form can be readily implemented in integrated photonic circuits, as shown in Fig.~\ref{fig:vbs}.

\subsection{Modified type-I fusion}

Using the same calculation techniques as in the previous section, we analyze the fusion of the same initial states \eqref{eqn:in_state_second_quantization}. For the modified type-I fusion gate, the successful outcomes split into two cases, each heralded by the detection of exactly one photon in either mode $1$ or $4$. Depending on the specific detection event, the two possible output states are:
    \begin{equation}\label{mod_Type_1_suc}
        \begin{split}
            |\psi^{(out)}\rangle_{1}& =  \frac{1}{\sqrt{P_1}}\left(\cos({\theta})\text{cos}(\alpha) \text{cos}(\beta)|0\rangle_c |\overline{0}\rangle_{A}|\overline{0}\rangle_{B} + \right.\\
            &\left. + \sin({\theta})\text{sin}(\alpha) \text{sin}(\beta) |1\rangle_c |\overline{1}\rangle_{A}|\overline{1}\rangle_{B}\right), \text{ or} \\
            |\psi^{(out)}\rangle_{2}& = \frac{1}{\sqrt{P_2}}\left(\sin({\theta})\text{cos}(\alpha) \text{cos}(\beta)|0\rangle_c |\overline{0}\rangle_{A}|\overline{0}\rangle_{B} - \right.\\
            &\left. - \cos({\theta})\text{sin}(\alpha) \text{sin}(\beta) |1\rangle_c |\overline{1}\rangle_{A}|\overline{1}\rangle_{B}\right),
        \end{split}
    \end{equation}
where corresponding success probabilities read
    \begin{equation}\label{mod_Type_1_prob}
        \begin{split}
            &P_1 =  \cos^2({\theta})\text{cos}^2 (\alpha) \text{cos}^2 (\beta) + \sin^2({\theta})\text{sin}^2 (\alpha) \text{sin}^2 (\beta), \\
            &P_2 =  \sin^2({\theta})\text{cos}^2 (\alpha) \text{cos}^2 (\beta) + \cos^2({\theta})\text{sin}^2 (\alpha) \text{sin}^2 (\beta).
        \end{split}
    \end{equation}
The total success probability $P_{s_1} = P_1+P_2$, coincides with the success probability of the standard fusion gate given in Eq.\eqref{Type_1_prob}. As evident from Eq.\eqref{mod_Type_1_suc}, the use of a variable beam splitter allows for control over the degree of entanglement in the output state through the parameter $\theta$.

In the case of failure, when either two or no photons are detected, the circuit yields a separable state $\hat{A}_0 \hat{B}_1 |vac\rrangle$ or $\hat{A}_1 \hat{B}_0 |vac\rrangle$.

\subsection{Modified type-II fusion}

Similarly, when fusing initial states of the form \eqref{eqn:in_state_second_quantization} using the gate depicted in Fig.\ref{fig:fusion_gates}(b), a successful operation is heralded by the detection of two photons of the input qubits in modes \{1,2\}, \{1,3\}, \{2,4\}, or \{3,4\}. The resulting output state depends on the specific detection pattern and takes the form:
\begin{equation}\label{mod_Type_2_suc}
    \begin{split}
        |\psi^{(out)}\rangle_{1,2}& =  \frac{1}{\sqrt{P_1}}\left(\pm\cos({\theta})\text{cos}(\alpha) \text{cos}(\beta)|\overline{0}\rangle_{A}|\overline{0}\rangle_{B} + \right.\\
        &\left. + \sin({\theta})\text{sin}(\alpha) \text{sin}(\beta) |\overline{1}\rangle_{A}|\overline{1}\rangle_{B}\right), \text{ or} \\
        |\psi^{(out)}\rangle_{3,4}& = \frac{1}{\sqrt{P_2}}\left(\pm\sin({\theta})\text{cos}(\alpha) \text{cos}(\beta)|\overline{0}\rangle_{A}|\overline{0}\rangle_{B} - \right.\\
        &\left. - \cos({\theta})\text{sin}(\alpha) \text{sin}(\beta)  |\overline{1}\rangle_{A}|\overline{1}\rangle_{B}\right).
    \end{split}
\end{equation}

The total success probability, $P_{s_1} = P_1+P_2$, matches the success probability of the standard fusion gate, shown in Fig.\ref{fig:succ_prob}. The modified type-II gates generate output states with entanglement properties similar to those of the type-I fusion gates \eqref{mod_Type_1_suc}. 

Fusion fails when both photons are detected in the same output mode or when detected in modes \{1,4\}. In such cases, the resulting state is separable and takes the form $\hat{A}_0 \hat{B}_1 |vac\rrangle$ or $\hat{A}_1 \hat{B}_0 |vac\rrangle$.

The success probability of the modified fusion of GHZ-like states can be boosted using the techniques described in \cite{Ewert2014} and \cite{Grice2011}, in a similar way as for the standard gates (see Section~\ref{sec:type2fusion}). The final states arising from additional successful measurement outcomes, however, exhibit degrees of entanglement different from those obtained in the unboosted cases (\ref{mod_Type_2_suc}).

\subsection{Fusing similar states}

In the special case where the input states $|G_{n}(\alpha)\rangle$ and $|G_{m}(\alpha)\rangle$  have similar degrees of entanglement, it is possible to deterministically set the desired Schmidt angle of the output state in the case of success using the following procedure:
\begin{enumerate}
    \item Apply single-qubit Pauli $X$ gates to all qubits of the first state, transforming it into $|G_{n}( \alpha)\rangle' = \text{sin}(\alpha) |\overline{0}\rangle + \text{cos}(\alpha)|\overline{1}\rangle$. In dual-rail encoding, this corresponds to a simple rearrangement of the modes of each qubit.
    \item Fuse the two states using a modified type-I fusion gate with parameter $\theta$. If exactly one photon is detected in either mode $1$ or $4$, the resulting state is:
    \begin{equation}
    \begin{split}
       &|\psi^{(out)}\rangle_{1} =\cos({\theta})|\overline{0}\rangle + \sin({\theta}) |\overline{1}\rangle  , \text{ or} \\
       &|\psi^{(out)}\rangle_{2} =\sin({\theta})|\overline{0}\rangle - \cos({\theta}) |\overline{1}\rangle.
    \end{split}
    \end{equation}   
    The total success probability is 
    \begin{equation} \label{eq:p_1}
        P_{s_2} = P_{1}+P_{2} = \sin^2({2\alpha})/2.
    \end{equation}
    \item If the photon is detected in mode 4, first apply a Pauli $Z$ gate to one qubit of the $|\psi^{(out)}\rangle_{(2)}$ state, then Pauli \(X\) gates to all qubits of the state.
\end{enumerate}

The output is a $|G_k(\theta)\rangle$ GHZ-like state of $k =(n+m-1)$ photons with a tunable entanglement degree $\theta$, determined by the chosen parameter of the modified fusion gate. This procedure enables the generation of output states with arbitrary entanglement. However, the success probability is limited to $P_{s_2} \leq 1/2$, in contrast to the general fusion success probability $P_{s_1} \leq 1$, defined by Eq.~\eqref{Type_1_prob}.

In the special case $\theta = \pi/4$, corresponding to a standard fusion gate, this procedure realizes entanglement distillation, yielding a maximally entangled GHZ state. The success probability \eqref{eq:p_1} matches that of a known distillation protocol for an ideal Bell-state measurement \cite{Kok2010}.

\subsection{Entanglement capability of the fusion gates}\label{entropy}

It is instructive to analyze the entangling capabilities of fusion gates using information-theoretic methods. The entanglement of a (generally mixed) state~$\rho$ can be characterized by the von Neumann entropy of the reduced density matrix~$\rho_A = \mathrm{Tr}_B(\rho)$~\cite{Kok2010}:

\begin{equation}
    S(\rho_A) \equiv -\text{Tr}(\rho_A \log_2 \rho_A).
\end{equation}

For a pure GHZ-like state for all splittings $A:B$ of the system the von Neumann entropy is given by:

\begin{equation}
    \begin{split}
        S(|G_n(\gamma)\rangle)= -\bigl(&\cos^2{(\gamma)}\log_2{\cos^2{(\gamma)}}+ \\
        +&\sin^2{(\gamma)}\log_2{\sin^2{(\gamma)}}\bigr).
    \end{split}
\end{equation}

To estimate the entangling power of fusion gates, we define the fusion entropy as the average entropy of the output states over all successful fusion outcomes:
\begin{equation}\label{eq:fus_ent}
    S_f=\sum_{i} P_i S(|G_n(\gamma_i)\rangle),
\end{equation}
where $P_i$ denotes the probability of the successful outcome $i$, and  $|G_n(\gamma_i)\rangle$ is the corresponding output state.

\begin{figure}[t]
    \includegraphics[width=3in]{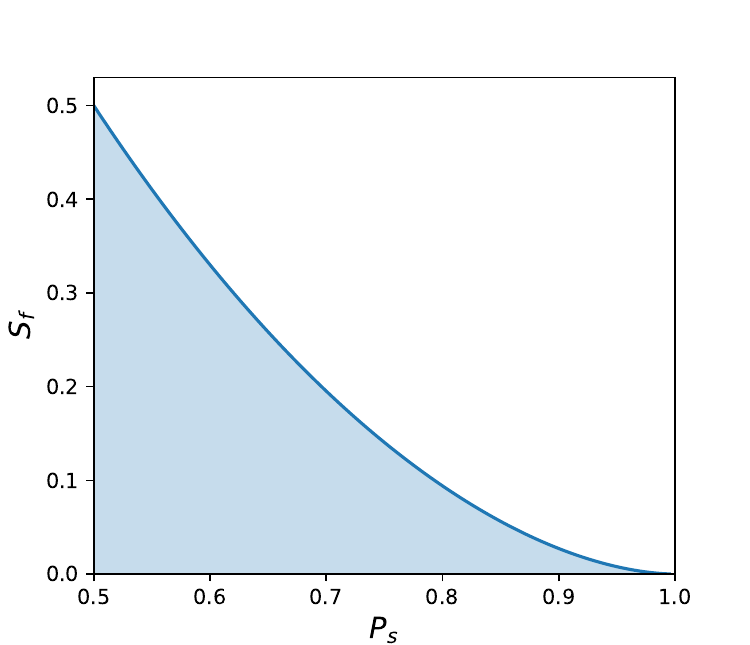}
    \caption{\textbf{Fusion entropy as a function of the total success probability} for the initial states with entanglement parameters $\alpha, \beta \in [0,\pi/4]$ using standard fusion gates. The solid line indicates the maximal entropy achievable for a given success probability.
    }
    \label{fig:entropy}
\end{figure}

Using this definition, we analyze the trade-off between fusion entropy and the total success probability for the fusion of GHZ-like states using standard fusion gates \eqref{Type_1_prob}, as shown in Fig.~\ref{fig:entropy}. The maximum entropy is achieved when the input states $|G_n(\alpha)\rangle$ and $|G_m(\beta)\rangle$ are maximally entangled, $\alpha=\beta=\pi/4$. For other values of 
$\alpha$ and $\beta$, although the fusion entropy is reduced, the success probability increases, and the initial states are less entangled, making them generally easier to prepare.

\begin{figure*}[ht]
    \includegraphics[width=7in]{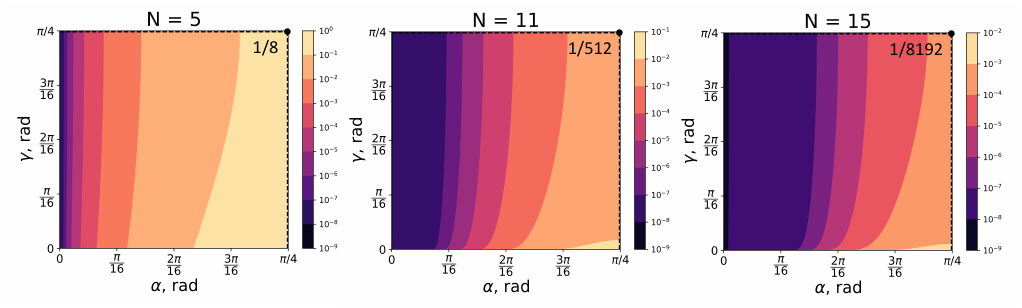}
    \caption{\textbf{Success probability of generating the GHZ-like state $|G_{N}(\gamma)\rangle$} using the scheme shown in Fig.~\ref{fig:first_scheme}, with $|G_{2}(\alpha)\rangle$ resource states. The points in the upper-right corners correspond to the special case $\alpha = \gamma = \pi/4$, representing the success probabilities of known fusion protocols for creating GHZ states from smaller maximally entangled states \cite{gimeno2016, Bartolucci2021}.}
    \label{fig:first_scheme_prob}
\end{figure*}

Compared to standard gates, modified fusion gates exhibit reduced entangling capability. This is evident, for example, in the fusion of two Bell pairs, and is also apparent for the entropy of the fusion (\ref{eq:fus_ent}) of arbitrary GHZ-like states using standard and modified fusion gates:
\begin{equation}
    \begin{split}
        &S_{f_{std}} - S_{f_{mod}} = P_{s1} (\log_2{P_{s1}}+\cos^2{\theta}\log_2{\cos^2{\theta}}+\\
        &+\sin^2{\theta}\log_2{\sin^2{\theta}}) - P_{1}\log_2{P_{1}} - P_{2}\log_2{P_{2}} \geq 0,
    \end{split}
\end{equation}
where $P_{s_1}$, $P_{1}$, $P_{2}$ are the outcome probabilities as defined in Eqs.~\eqref{Type_1_prob} and \eqref{mod_Type_1_prob}, and $\theta$ is a tunable parameter of the modified gate. Consequently, modified fusion gates are particularly useful in applications that require adjustable entanglement degrees in the output states.

\section{Fusion-based state generation}\label{sec:state_gen}

\subsection{General method}

\begin{figure}[h]
    \includegraphics[width=3.3in]{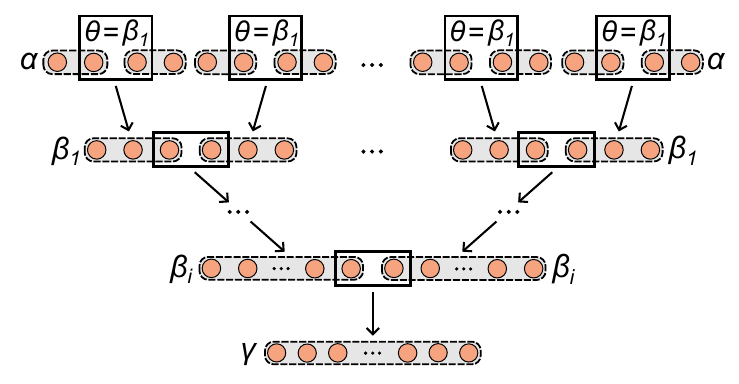}
    \caption{\textbf{Step-by-step scheme for generating an arbitrary GHZ-like state $|G_{N}(\gamma)\rangle$}. The protocol uses $2$-qubit resource states $|G_2(\alpha)\rangle$, which are first fused using modified fusion gates (rectangles labeled with angle $\theta$), followed by standard fusion gates (empty rectangles). All standard fusion operations can be applied in arbitrary order.
    }
    \label{fig:first_scheme}
\end{figure}

Large GHZ-like states can be generated by preparing multiple small resource states and fusing them using the procedures proposed in Sections \ref{sec:fusion_gates} and \ref{sec:mod_f_g}. In particular, an arbitrary target GHZ-like state $|G_{N}(\gamma)\rangle$ can be constructed from $N-1$ resource states $|G_{2}(\alpha)\rangle$ via the following steps, illustrated in Fig.~\ref{fig:first_scheme}:
\begin{enumerate}
    \item Fuse the initial resource states pairwise using modified type-I fusion gates with parameter $\theta = \beta_1$. This yields $(N-1)/2$ intermediate states $|G_3(\beta_1)\rangle$, where the entanglement degree $\beta_1 = \arctan\bigl(\tan^{2/(N-1)}({\gamma})\bigr)$.
    \item Fuse the resulting states together using standard type-I fusion gates.
\end{enumerate}

This procedure results in the target GHZ-like state $|G_{N}(\gamma)\rangle$, with a total success probability given by
\begin{equation}\label{eq:first_scheme_prob}
    P_{\text{gen}}^{(1)}(\alpha, \gamma) = \bigl(P_{s_2}(\alpha)\bigr)^{\frac{N-1}{2}}\prod P_{s_1},
\end{equation}
where $P_{s_1}$ and $P_{s_2}$ are the success probabilities of the fusion procedures \eqref{Type_1_prob} and \eqref{eq:p_1} respectively. The second product runs over the $(N-3)/2$ standard fusion operations required in the second stage. 

The total success probability of the scheme for different numbers of qubits is shown in Fig.~\ref{fig:first_scheme_prob}. 

The scheme is designed to minimize the use of modified fusion gates, which exhibit lower entangling capability than standard gates, as shown in Section~\ref{entropy}.

If $N$ is even, two of the initial $2$-qubit resource states should be fused using a type-II fusion gate, resulting in one $|G_2(\beta_1)\rangle$ state at the first stage.

In the special case $\alpha=\gamma=\pi/4$, the scheme reduces to previously known protocols for constructing maximally entangled GHZ states from smaller maximally entangled resource states \cite{Bartolucci2021, gimeno2016}.

The type-I gates used in this method do not allow for loss tracking, unlike type-II gates, which measure all modes of the input qubits. However, we assume that losses occurring during type-I fusion operations can be heralded at later stages of the protocols that utilize GHZ-like states generated by the proposed scheme. For instance, in a quantum computing framework similar to FBQC~\cite{Bartolucci2023}, such losses arising during the state-generation process can be accounted for in conjunction with losses occurring in type-II fusions during the computational stage of the FBQC protocol.

\subsection{More efficient method}\label{sec:target_nghz}
The scheme presented above enables the creation of a target GHZ-like state from a set of similar arbitrary-entangled resource states. However, its overall success probability can be significantly improved by using specifically tailored resource states.

\begin{figure}[h]
    \includegraphics[width=3.3in]{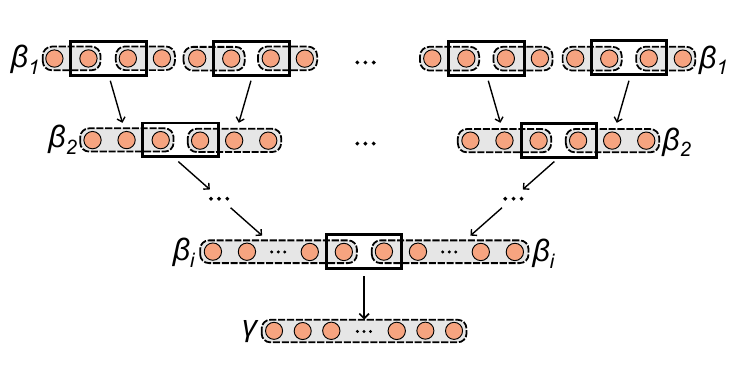}
    \caption{\textbf{More efficient procedure for generation of GHZ-like states $|G_{N}(\gamma)\rangle$.}  The procedure exploits $2$-qubit resource states  $|G_2(\beta_1)\rangle$ and standard type-I fusion gates, with $\beta_1 = \arctan\bigl(\tan^{1/(N-1)}(\gamma)\bigr)$.}
    \label{fig:second_scheme}
\end{figure}

A target $N$-qubit GHZ-like state $|G_{N}(\gamma)\rangle$ can be created from  $(N-1)$ two-qubit resource states $|G_2(\beta_1)\rangle$ with the Schmidt angle $\beta_1 = \arctan\bigl(\tan^{1/(N-1)}(\gamma)\bigr)$, by fusing them using standard fusion gates, as illustrated in Fig.~\ref{fig:second_scheme}.

\begin{figure}[ht]
        \includegraphics[width=3.3in]{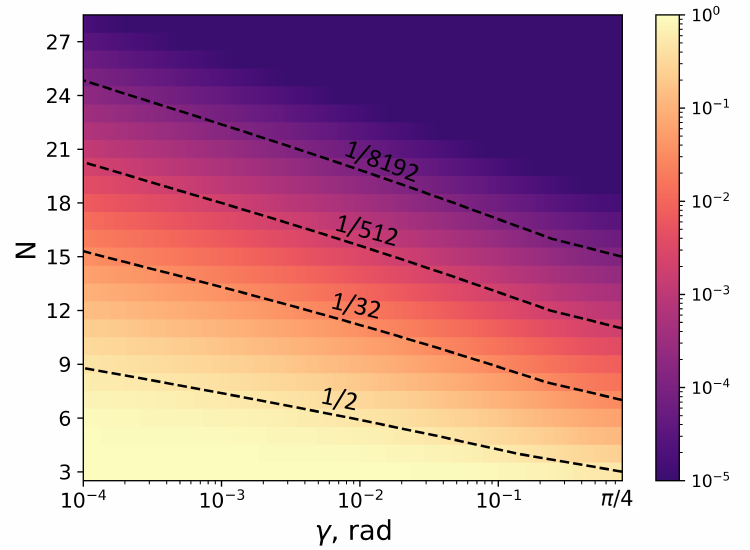}
        \caption{\textbf{Success probability for generating the GHZ-like state $|G_{N}(\gamma)\rangle$} by fusing two-qubit resource states $|G_2(\beta_1)\rangle$ using standard fusion gates. The resource state Schmidt angle is chosen as $\beta_1 = \arctan\bigl(\tan^{1/(N-1)}(\gamma)\bigr)$, ensuring the desired entanglement in the output. For $\gamma = \pi/4$, the scheme reduces to known protocols for generating maximally entangled GHZ states \cite{Bartolucci2021, gimeno2016}.}
        \label{fig:second_scheme_prob}
    \end{figure}

The total success probability of this scheme is given by:
\begin{equation}\label{eq:second_scheme_prob} P_{\text{gen}}^{(2)}(\gamma) = \prod P_{s_1}\bigl(\beta_i, \beta_j\bigr), 
\end{equation} 
where the product runs over all fusion operations in the scheme. The success probability as a function of the target state's parameters $N$ and $\gamma$ is shown in Fig.~\ref{fig:second_scheme_prob}.

This method corresponds to the second stage of the general scheme shown in Fig.\ref{fig:first_scheme}, assuming the resource states $|G_2(\beta_1)\rangle$ being generated at the beginning. In contrast to the general method, every fusion in Eq.\eqref{eq:second_scheme_prob} succeeds with a probability $P_{s_2}\geq 1/2$ (\ref{Type_1_prob}), resulting in a higher overall success probability.

In the special case of maximally entangled target states $|G_{N}(\pi/4)\rangle$, the required resource states are $|G_2(\pi/4)\rangle$, and the scheme aligns with previously known protocols for generating GHZ states from smaller maximally entangled states \cite{Bartolucci2021, gimeno2016}. The corresponding success probabilities are shown in Fig.~\ref{fig:second_scheme_prob} for $\gamma=\pi/4$.

Although this second, more efficient method offers a higher overall success probability, it requires specific resource states. In contrast, the general method can operate with arbitrarily entangled resource states. Both methods can be readily adapted to work with input states of varying qubit numbers.

\section{Resource requirements}\label{sec:performance}

\subsection{Creation of small resource states}

We have proposed schemes for generating arbitrary GHZ-like target states using small entangled resource states and probabilistic entangling measurements. These resource states can be prepared by transforming single-photon inputs using multimode interferometers, followed by projective measurements on a subset of modes \cite{Forbes2025}.

Several proposals have introduced methods for the probabilistic generation of photonic dual-rail encoded Bell states \cite{Carolan2015, Bartolucci2021, Fldzhyan2021, gimeno2016} and GHZ states \cite{Bartolucci2021, Varnava2008, Gubarev2020, Bhatti2024, gimeno2016} from single photons.

Recent works \cite{Fldzhyan2023, Skryabin2024} have proposed circuits capable of the tunable generation of arbitrary two-qubit GHZ-like states of the form $\cos(\alpha)|00\rangle + \sin(\alpha)|11\rangle$ from four single photons. The efficiency of two-qubit state generation in these schemes depends slightly on the degree of entanglement, being higher for states with lower entanglement.

If the two-qubit resource states $|G_{2}(\alpha)\rangle$ or $|G_{2}(\beta_1)\rangle$ used in our schemes are generated at a rate $f_r$, then the rate of generating $N$-qubit GHZ-like states is given by
    \begin{equation}
        f_t=\frac{f_r}{N-1} P_{\text{gen}},
    \end{equation}
where $P_{\text{gen}}$ is the total success probability defined in Eqs.~(\ref{eq:first_scheme_prob}) and (\ref{eq:second_scheme_prob}) for the first and second schemes, respectively.

\begin{figure}[t]
        \includegraphics[width=3.15in]{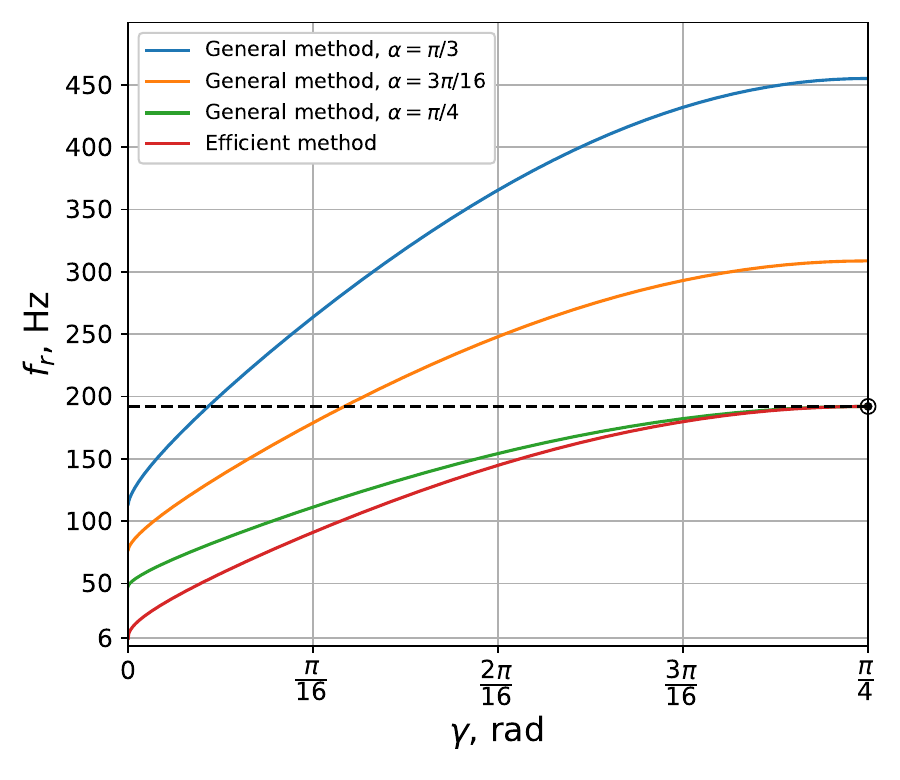}
        \caption{\textbf{Resource state generation rate required to obtain the $7$-qubit GHZ-like states at a target rate of $f_t=1$ Hz.} The dependence of the generation rate on the target state entanglement degree. The dashed line shows the required Bell pair generation rate to produce maximally entangled states $|G_{7}(\pi/4)\rangle$ at the same target rate using known fusion-based methods \cite{Bartolucci2021, gimeno2016}.}
        \label{fig:comparison}
    \end{figure}

Fig.~\ref{fig:comparison} estimates the required resource state generation rate $f_r$ to achieve a target generation rate $f_t=1 \; \text{Hz}$ for a $7$-qubit GHZ-like state. In the special case $\alpha=\gamma =\pi/4$, both methods reduce to previously proposed fusion schemes for generating maximally entangled GHZ states from maximally entangled resource states \cite{Bartolucci2021, gimeno2016}. The dashed line in Fig.~\ref{fig:comparison} illustrates the performance of such a scheme using six Bell pairs to generate $|G_{7}(\pi/4)\rangle$ state, requiring a resource generation rate of $f_r=(N-1)2^{N-2} \approx 192 \; \text{Hz}$.

Recent experiments have already demonstrated heralded two-qubit resource state generation rates of $f_r=6$ Hz \cite{Li2021} and $f_r=2.36$ Hz \cite{Skryabin2024} for Bell and $2$-qubit GHZ-like states, respectively. These results, with considerable potential for further improvement, highlight the promise of linear-optical entanglement generation for applications in quantum computing and communication.

\subsection{Multiplexing}

Practical linear-optical applications require multiplexing schemes to achieve near-deterministic entanglement generation. In this section, we estimate the average number of single photons required to deterministically generate a GHZ-like state using multiplexing for the efficient method, and compare it to established techniques for creating maximally entangled GHZ states, described in \cite{Bartolucci2021}.

The concept of perfectly resource-efficient multiplexing \cite{Bartolucci2021} assumes that at each stage of the protocol, each of $M$ probabilistic sources has a success probability $p_0$, and all successful result are proceeded to the further stages. In the limit $M \rightarrow \infty$, this yields $M{}p_0$ successful outcomes per stage. If each source consumes $n_0$ single photons, then the average number of single photons required for one successful output at that stage is
    \begin{equation}
        \overline{n} = \frac{M{}n_0}{M{}p_0} = \frac{n_0}{p_0}.
    \end{equation}

Further we assume that all resource states $|G_r(\beta_1)\rangle$ are generated from single photons with the same probabilities as their maximally entangled counterparts $|G_r(\pi/4)\rangle$. Following \cite{gimeno2016}, we take that the two-qubit resource states can be generated from $4$ single photons with a success probability $p_{succ}=1/8$, and three-qubit resource states can be generated from 6 single photons with a success probability $p_{succ}=1/32$. 

Fig.~\ref{fig:num_photons} presents the estimated average number of single photons required to deterministically generate target GHZ-like states using different configurations of resource states for the efficient method from Section~\ref{sec:target_nghz}:

\begin{figure}[t]
    \includegraphics[width=3.3in]{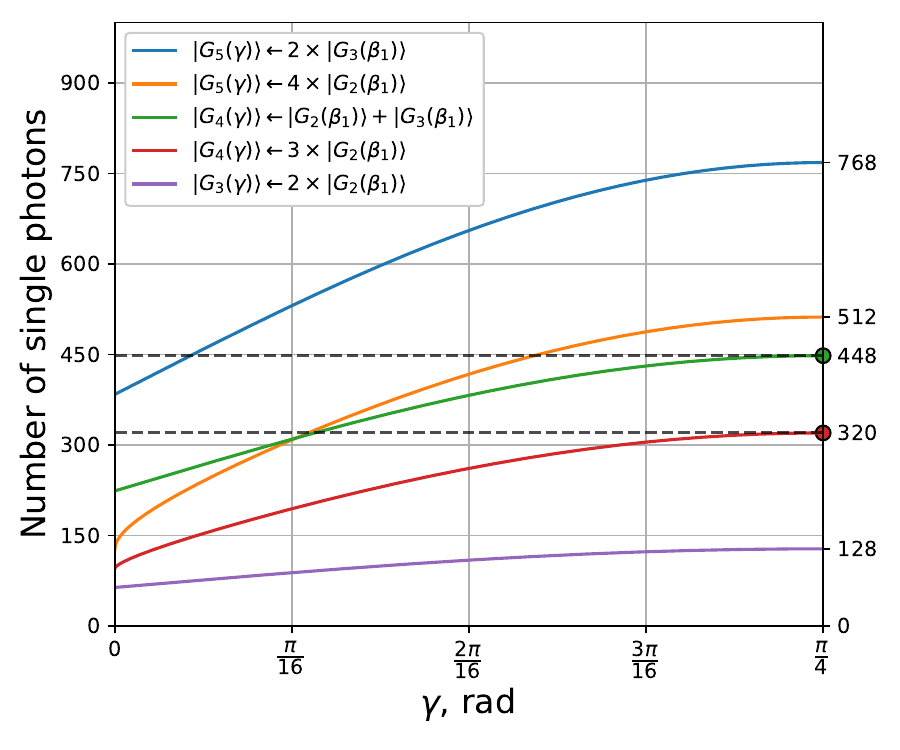}
    \caption{\textbf{Average number of single photons required for near-deterministic generation of target GHZ-like states $|G_{N}(\gamma)\rangle$ using various configurations of initial states in the efficient method.} Legend shows target states and corresponding initial state configurations. Data points on the right correspond to the known fusion schemes for generating maximally entangled GHZ states ($\gamma = \pi/4$), established in \cite{Bartolucci2021}.}
    \label{fig:num_photons}
\end{figure}

\begin{enumerate}
    \item 5-qubit $|G_5(\gamma) \rangle$ target state from two $3$-qubit resource states $|G_3(\beta_1) \rangle$ (12 single photons in total), $\beta_1 = \arctan\bigl(\tan^{1/2}(\gamma)\bigr)$.

    \item 5-qubit $|G_5(\gamma) \rangle$ target state from four $2$-qubit states $|G_2(\beta_1) \rangle$ (16 single photons in total), $\beta_1 = \arctan\bigl(\tan^{1/4}(\gamma)\bigr)$.

    \item 4-qubit $|G_4(\gamma) \rangle$ target state from a $2$-qubit $|G_2(\beta_1) \rangle$ and a $3$-qubit $|G_3(\beta_1) \rangle$ resource states (10 single photons in total), $\beta_1 = \arctan\bigl(\tan^{1/2}(\gamma)\bigr)$.

    \item 4-qubit $|G_4(\gamma) \rangle$ target state from three $2$-qubit states $|G_2(\beta_1) \rangle$ (12 single photons in total), $\beta_1 = \arctan\bigl(\tan^{1/3}(\gamma)\bigr)$.

    \item 3-qubit $|G_3(\gamma) \rangle$ target state from two $2$-qubit resource states $|G_2(\beta_1) \rangle$ (8 single photons in total), $\beta_1 = \arctan\bigl(\tan^{1/2}(\gamma)\bigr)$.

\end{enumerate}

Note that these results represent lower bounds on the performance of the schemes since non-maximally entangled states $|G_r(\beta_1) \rangle$ are typically easier to generate than their maximally entangled counterparts \cite{Fldzhyan2023}.

Fig.~\ref{fig:num_photons} provides a comparison to the methods for creating maximally-entangled states proposed in \cite{Bartolucci2021}. Another scheme creating $|G_4(\pi/4) \rangle$ via the so-called non-qubit primate resource states was also proposed in \cite{Bartolucci2021}, requiring approximately 309 single photons. The potential of employing primate states for the generation of GHZ-like states was investigated in a recent work \cite{Fldzhyan2025}.

Fig.~\ref{fig:comparison} and \ref{fig:num_photons} demonstrate that our methods may create GHZ-like states with significantly higher efficiencies than existing techniques for producing maximally entangled GHZ states.

\section{Conclusion}\label{sec:conclusion}

In this work, we have investigated ways of creating and manipulating GHZ-like states, a class of states with varying degrees of entanglement, including standard GHZ states as a special case. We have shown that these states can be generated and fused with significantly higher efficiency than maximally entangled GHZ states, though at the cost of reduced entanglement in the resulting states. GHZ-like states are known to be useful in quantum teleportation protocols \cite{Jeong2006, Modawska2008, PATHAK2011}, and are local-unitary equivalent to weighted hypergraph states \cite{https://doi.org/10.48550/arxiv.2408.02740}, which are encompassed within recent extensions of the Pauli stabilizer formalism \cite{Webster2022}.

The first proposed scheme enables the creation of target GHZ-like states using arbitrary entangled resource states. The second scheme achieves higher generation efficiency, but requires resource states with specific entanglement degrees. Both approaches benefit from the entangling properties of fusion measurements, and may be further enhanced through alternative configurations of resource states and fusion gates. The performance of both schemes can also be improved using boosting techniques \cite{Grice2011, Ewert2014, https://doi.org/10.48550/arxiv.2410.20261}, though these require additional photonic resources.

The creation of entangled photonic states remains one of the central challenges for linear-optical quantum technologies. Current methods for generating dual-rail photonic stabilizer states using only linear optical elements achieve relatively low success probabilities \cite{gimeno2016, Bartolucci2021}. Alternative approaches involve application of quantum emitters \cite{Chan2025}, such as quantum dots \cite{Lindner2009} or single atoms \cite{Thomas2022}, which enable on-demand photonic state generation and benefit from advances in corresponding technologies \cite{Shi2024, Cogan2018}.

Our analysis shows that GHZ-like states with tunable degrees of entanglement can be generated using linear-optics with significantly higher success probabilities than standard GHZ states. Moreover, we demonstrate that such states can be fused with probabilities exceeding 50\% without using additional photonic resources, which is crucial for efficient realization of fusion-based quantum computing \cite{Bartolucci2023} or percolation-based protocols \cite{GimenoSegovia2015, Pant2019}. The ability to finely tune the degrees of entanglement of GHZ-like states through modified fusion gates also provides a versatile tool for controlling such states.

With recent advances in the study of quantum states beyond the stabilizer formalism \cite{https://doi.org/10.48550/arxiv.2408.02740, Webster2022, Gross2007, Ni2015}, GHZ-like states appear to be highly promising candidates for resources for future linear-optical quantum applications.

\section{Acknowledgements}

A. Melkozerov is grateful to the Russian Foundation for the Advancement of Theoretical Physics and Mathematics (BASIS) (Project №24-2-2-5-1). 

\bibliographystyle{unsrt}

\end{document}